\documentclass[12pt]{article}
\usepackage[margin=2.5cm]{geometry}
\usepackage{amsmath,amssymb}
\usepackage{graphicx}
\usepackage{bm}
\usepackage{cite}
\usepackage{xcolor}
\begin{document}

\begin{center}
{\Large\bf Magnetic Modulation of Ionic Transport in a Multistate Channel Model}\\[4mm]
Tina Arabi and Ehsan Noruzifar\\[2mm]
{\small Department of Physics, K. N. Toosi University of Technology, P.O. Box 15875-4416, Tehran, Iran\\
E-mail: noruzifar@kntu.ac.ir}
\end{center}

\begin{abstract}
We study the magnetic modulation of ionic transport in an idealized multistate channel model. The channel can occupy 
different structural states, each with its own ionic current and magnetic response. We assume quasi-equilibrium Boltzmann 
populations for the structural states, while ionic transport is driven by an externally imposed electrochemical potential 
difference. Using a statistical-mechanical linear-response approach, we relate changes in the average current to magnetic 
perturbations of the structural states. The response is determined by the covariance between the state-dependent ionic 
currents and their magnetic-response coefficients, which controls both its sign and magnitude. We also examine the magnetic 
and thermal sensitivities, illustrate the results with a three-state model. Order-of-magnitude estimates suggest that the 
predicted effect for a single channel is extremely small at millitesla magnetic fields. Overall, the results provide a 
quantitative criterion for assessing magnetic modulation of transport within this idealized model and may help in 
considering possible effects in real ion channels and other multistate transport systems.
\end{abstract}


\section{Introduction}

Ion channels are membrane proteins that conduct ions across cell membranes through nanometre-scale pores and can switch between different conformational substates on a rugged free-energy landscape \cite{Frauenfelder1991, Hille2001}. Their stochastic gating has been studied for many years from the perspective of statistical physics. Different approaches include kinetic models, conformational-diffusion models, and path-probability descriptions of gating \cite{Braun1998, Goychuk2003, Ozer2004}, as well as studies of fluctuations, flicker noise, and collective effects in ensembles of channels \cite{Schmid2001, Kosinska2008, Lee1999, Lee2002} and of the stochastic thermodynamics of a probe in a fluctuating correlated field \cite{Venturelli2024}. Statistical-mechanical models have also considered the direct coupling of an external field to channel populations \cite{Yang1999}. Ionic transport through the pore has likewise been described at different levels. These range from Markovian models of K$^+$ permeation through the KcsA selectivity filter, which predict a current consistent with the measured single-channel value of about $11~\text{pA}$ \cite{KcsA2010}, and Brownian-dynamics and Monte Carlo models of the selectivity filters of calcium channels \cite{Kaufman2013, Giri2011}, to continuum Poisson--Nernst--Planck models of electrodiffusion, including steric, self-energy-modified, and electroneutral versions \cite{Bazant2007, Xu2014, Song2018}, together with the computation of the electric field in proteins and channels \cite{Eisenberg1996}. Reviews of permeation models cover continuum theories, Brownian dynamics, and molecular dynamics \cite{Kuyucak2001}. Reduced Fick--Jacobs-type models describe diffusion in narrow channels \cite{Biharmonic2023}, diffusive exit through narrow pores has been analyzed \cite{Yang2023}, and noise-induced transport has been studied in deformable tubes \cite{Deformable2012}. Fast and selective ionic transport has also been compared between biological channels and synthetic ion-exchange membranes \cite{Kreuer2021}. In these studies, permeation is considered for a given pore geometry. In the present work, the geometry associated with each conformational substate is assumed to remain fixed, while the magnetic field changes the populations of the substates. In all such descriptions, the macroscopic current can be viewed as an average of the currents associated with individual states, weighted by their conformational populations. Thus, the current carried by a state and the probability of occupying that state represent two separate ingredients. A weak perturbation that changes the relative free energies of the states may therefore also change the average current.
 
The control of state populations by an external parameter is well known. An external electric field applied to a channel can change the fraction of channels in the open state \cite{Yang1999}. Fluctuating voltages can also focus voltage-gated channels into particular conformational states: in experiments on Shaker K$^+$ channels, dichotomous noise increased the probability of intermediate states by up to about $50\%$ above the maximal value reached under any static voltage \cite{Kargol2008}. In electroconformational-coupling theory, an applied field induces conformational transitions in a membrane protein, and both signal and noise influence the resulting transport \cite{Tsong2003}; noise-driven transport on the nanoscale has been reviewed for artificial Brownian motors \cite{Hanggi2009}. Temperature can affect both the gating kinetics and the equilibrium populations. For example, path-probability models of voltage-gated channels show that increasing temperature shifts the distribution of the gating time constant toward longer times \cite{Ozer2004}. A related example of population redistribution is provided by the heat-activated channel TRPV1. Its open probability is controlled by a large enthalpy difference between the closed and open states \cite{Yao2011}, and a temperature change of only a few kelvin can substantially redistribute the populations. A static magnetic field is expected to be a much weaker control parameter. The question considered here is whether it can influence the populations through a similar mechanism and, if so, how this influence appears in ionic transport.
 
The effects of magnetic fields on ion channels and membrane transport have remained controversial for a long time. Experimental studies of voltage-activated channels, calcium channels, and cell membranes have not led to a clear conclusion \cite{Zhadin1990, Zhadin2001, Hojevik95, Obo2002, Koch2003, Rosen2003, Fan2012}. A systematic review of the effects of electromagnetic fields on neuronal ion channels reports diverse effects that depend on the field frequency, the exposure time, and the properties of the tissue, with changes in calcium homeostasis being the most commonly reported result \cite{Bertagna2021}. In a recent patch-clamp study, exposure to a $1~\text{mT}$, $50~\text{Hz}$ field decreased inward and transient outward currents in hippocampal neurons, and this effect was attributed to altered calcium homeostasis \cite{Bertagna2025}. Theoretical explanations based on Lorentz forces or ion-cyclotron resonance \cite{Liboff1988} have been questioned on physical grounds \cite{Halle1988, Sandweiss1990}. In addition, phenomenological models for magnetic effects on calcium signaling \cite{Zhang2009} do not explicitly connect the response to the conformational energetics of the channel protein. A statistical-mechanical criterion relating the magnetic energetics of the protein to the currents associated with its individual conformational states is still lacking.
 
Such a criterion can be obtained by combining two established ideas. First, closed-shell proteins are diamagnetic, and the peptide groups in ordered $\alpha$-helices possess an anisotropic magnetic susceptibility \cite{Pauling1979, Worcester1978, Torbet1981, Maret1986}. As a result, a magnetic field can change the free energy of a conformation by an amount proportional to $B^2$, with a dependence on the orientation of the field \cite{Landau1984}. Second, under a weak perturbation of Boltzmann populations, the response of an observable can be expressed as a covariance between the observable and the perturbing energy \cite{Kubo1966}. Taken together, these two ideas lead to the general form of the magnetic response in a multistate channel.
 
In this work, we apply this idea to a coarse-grained multistate channel model. The conformational populations are assumed to be in quasi-equilibrium, whereas the ionic flux is driven by an imposed electrochemical potential difference. Nonequilibrium fluctuations in ionic transport through biomembranes have been studied previously \cite{Lee1999, Lee2002}; in the present model, as a first approximation, conformational relaxation is taken to be fast compared with the observation time, allowing the populations to be described by a Boltzmann distribution. The main results can be summarized as follows: (i)~The sign and magnitude of the magnetic modulation of the average current are determined by the covariance between the currents of the individual states and their state-dependent magnetic response coefficients. The response also has a $\cos^2\theta$ dependence on the field orientation. (ii)~The magnetic and thermal sensitivity parameters, $S_B$ and $S_T$, arise from the same covariance structure. While $S_B$ increases linearly with $B$, $S_T$ is determined by the covariance between the current and the energy. (iii)~Cauchy--Schwarz inequalities provide bounds on the response in terms of fluctuations of the state currents and the structural order parameter. For a single channel in millitesla magnetic fields, the estimated relative modulation is of order $10^{-9}$. This is consistent with the general expectation that static magnetic energies at such field strengths are much smaller than $k_{\rm B}T$. The purpose of this work is therefore not to predict a large magnetic effect, but to obtain an explicit covariance structure and the corresponding bounds. (iv)~A three-state example, with parameters based on TRPV1, illustrates the behavior of these quantities over the physiological temperature range. The model is idealized, and its limitations, especially the assumed separation between structural populations and ionic transport, are discussed in Section~\ref{sec:discussion}.

\section{Statistical-Mechanical Framework}
\label{sec:model}

We consider a coarse-grained multistate ion channel embedded in a membrane and in contact with a thermal reservoir at temperature $T$. The channel is also 
exposed to a stationary electrochemical driving force $V$ and an external static magnetic field $\mathbf{B}$. At the conformational level, the protein can 
occupy a discrete number of $N$ metastable structural states, which are referred to as conformational substates and are indexed by $i \in 
{1,\dots,N}$~\cite{Frauenfelder1991,Hille2001}. Each substate $i$ has its own conformational free energy $\mathcal{F}_i(\mathbf{B}, T)$ and is associated 
with a conditional stationary ionic current $I_i(V)$.

The definition of a state-resolved current $I_i(V)$ requires an appropriate separation of time scales. We assume that the mean residence time of the 
channel in a given conformational state $i$ is much longer than the microscopic transit time of individual ions through the pore. This allows the ionic 
current associated with each state to reach a stationary value before the channel changes its conformation. At the same time, the mean residence time is 
assumed to be much shorter than the macroscopic observation time, so that the measured current includes contributions from different conformational 
states. With this assumption, averaging over conformational fluctuations gives the stationary mean current,
\begin{equation}
I(\mathbf{B}, T, V) = \sum_{i=1}^{N} p_i(\mathbf{B}, T), I_i(V),
\label{eq:current}
\end{equation}
where $p_i(\mathbf{B}, T)$ is the stationary probability of finding the channel in substate $i$ under the applied magnetic field and at temperature $T$.

In order to study the statistical-mechanical contribution to magnetic and thermal modulation without introducing specific kinetic transition rates, we use a quasi-equilibrium description of the structural states. In this description, the coupling between conformational dynamics and ionic conduction is considered in one direction only. More specifically, the ionic flux within each open or partially conducting substate is a driven nonequilibrium process determined by the applied electrochemical potential difference $V$. However, the conformational transitions are assumed to reach thermal equilibrium on the structural free-energy landscape ${\mathcal{F}_i(\mathbf{B}, T)}$. This means that the conformational populations are not affected by ionic occupancy or by feedback from electrodiffusion. In particular, both $\mathcal{F}_i$ and $p_i$ are assumed to be independent of $V$. Under these conditions, the structural ensemble can be described by the canonical partition function
\begin{equation}
\mathcal{Z}(\mathbf{B}, T) = \sum_{i=1}^{N} \exp\big[-\beta \mathcal{F}_i(\mathbf{B}, T)\big],
\label{eq:partition}
\end{equation}
where $\beta = (k_{\mathrm{B}} T)^{-1}$ and $k_{\mathrm{B}}$ is the Boltzmann constant. The corresponding occupation probability of each conformational substate is then given by
\begin{equation}
p_i(\mathbf{B}, T) = \frac{\exp\big[-\beta \mathcal{F}_i(\mathbf{B}, T)\big]}{\mathcal{Z}(\mathbf{B}, T)}.
\label{eq:Boltzmann}
\end{equation}

In the following, angular brackets denote ensemble averages calculated using the unperturbed Boltzmann distribution at zero magnetic field, $p_i^{(0)} = 
p_i(\mathbf{0}, T)$. Thus, for any state-dependent quantity $A$, its average is defined as $\langle A \rangle = \sum_i p_i^{(0)} A_i$. We also introduce 
the equilibrium covariance between two state-dependent quantities $A$ and $X$ as
\begin{equation}
\mathrm{Cov}(A, X) \equiv C_{AX} = \langle A X \rangle - \langle A \rangle \langle X \rangle.
\label{eq:cov_def}
\end{equation}

We now assume that the applied magnetic field produces a change in the conformational free energy of each state. This change can be written as
\begin{equation}
\mathcal{F}_i(\mathbf{B}, T) = \mathcal{F}_i^{(0)}(T) + \Delta\mathcal{F}_i(\mathbf{B}),
\end{equation}
where $\mathcal{F}_i^{(0)}(T)$ is the free energy in the absence of a magnetic field and $\Delta\mathcal{F}_i(\mathbf{B})$ is the field-induced 
contribution. Such a change can be viewed as a modification of the conformational free-energy landscape, which may favor some structural states over 
others~\cite{Sigg2014,Delemotte2015}. For biological macromolecules exposed to non-destructive magnetic fields under laboratory or environmental 
conditions, the magnetic interaction energy is expected to be much smaller than the thermal energy. Therefore, we have $|\Delta\mathcal{F}i| \ll 
k_{\mathrm{B}} T$ for all thermally accessible substates. This condition allows us to expand the partition function~\eqref{eq:partition} and the occupation 
probabilities~\eqref{eq:Boltzmann} to first order in $\beta \Delta\mathcal{F}_i$. The resulting change in the population of each state is
\begin{equation}
\Delta p_i = p_i(\mathbf{B}, T) - p_i^{(0)} = -\beta\, p_i^{(0)} \big(\Delta\mathcal{F}_i - \langle \Delta\mathcal{F} \rangle\big) + 
\mathcal{O}\left((\beta \Delta\mathcal{F})^2\right).
\label{eq:deltap}
\end{equation}

This expression shows how a weak magnetic perturbation changes the relative populations of the conformational states. States with a magnetic free-energy 
shift smaller than the ensemble average $\langle \Delta\mathcal{F} \rangle$ become more populated, whereas states with a larger shift become less 
populated. The magnitude of this redistribution is controlled by the ratio of the magnetic energy change to the thermal energy, with thermal fluctuations 
tending to reduce the effect of the perturbation. In this way, Eq.~\eqref{eq:deltap} describes the competition between the magnetic modification of the 
conformational landscape and thermal fluctuations.

By substituting Eq.~\eqref{eq:deltap} into Eq.~\eqref{eq:current}, 
and assuming that the pore geometry does not change within a given conformational substate, so that 
$\partial I_i / \partial B \approx 0$, we obtain the relation between the magnetic perturbation and the change in the average current. Defining the 
field-induced current modulation as $\Delta I(\mathbf{B}, T, V) = I(\mathbf{B}, T, V) - I(\mathbf{0}, T, V)$, we find
\begin{equation}
\Delta I = -\beta \sum_{i=1}^{N} p_i^{(0)} \big(I_i - \langle I \rangle\big) \Delta\mathcal{F}_i = -\beta\, \mathrm{Cov}(I, \Delta\mathcal{F}),
\label{eq:MasterResponse}
\end{equation}
which is valid to leading order in $\beta \Delta\mathcal{F}$. Equation~\eqref{eq:MasterResponse} 
is a direct result of canonical linear response theory~\cite{Kubo1966}. 
It shows that the change in the macroscopic current due to a weak magnetic perturbation is determined by the equilibrium covariance, evaluated at zero 
magnetic field, between the state-dependent ionic currents and the corresponding magnetic free-energy shifts. In other words, the magnetic response 
depends not only on the magnitude of the free-energy changes, but also on how these changes are correlated with the currents carried by the different 
conformational states. This result also gives a simple condition for the absence of magnetic modulation. If the ionic current (or equivalently the 
conductance at a fixed driving force) is the same in all substates, or if the magnetic free-energy shift is identical for all of them, the covariance 
becomes zero. In either case, the magnetic field does not produce a change in the mean current within the present model, irrespective of the magnitude of 
the applied field.

\section{Magneto-Thermal Response and Sensitivity}
\label{sec:quadratic}

For a closed-shell diamagnetic protein assembly without intrinsic ferromagnetic ordering or 
persistent spin-radical pairs, time-reversal invariance requires the conformational free energy of each 
equilibrium substate to be an even function of the magnetic field, 
$\mathcal{F}_i(\mathbf{B}) = \mathcal{F}_i(-\mathbf{B})$~\cite{Landau1984}. 
Therefore, terms linear in $\mathbf{B}$ are forbidden, 
and the leading magnetic contribution has the following quadratic tensorial form:
\begin{equation}
\Delta\mathcal{F}_i(\mathbf{B}) = -\sum_{a,b} B_a\, c_{ab}^{(i)}\, B_b + \mathcal{O}(B^4),
\label{eq:dFgeneral}
\end{equation}
where $a,b \in \{x,y,z\}$ are Cartesian indices and $c_{ab}^{(i)} = c_{ba}^{(i)}$ is the 
symmetric second-rank magnetic-response tensor of substate $i$. With the minus sign convention 
used in Eq.~\eqref{eq:dFgeneral}, a positive response component decreases the conformational free energy as the magnetic field increases.

Ion channels generally have approximate rotational or $C_n$ axial symmetry around the transmembrane pore axis, 
represented here by the unit vector $\mathbf{u}$. Using an effective uniaxial description for the conformational ensemble, 
the tensor $c_{ab}^{(i)}$ can be separated into isotropic and uniaxial-anisotropic contributions:
\begin{equation}
c_{ab}^{(i)} = a_i\, \delta_{ab} + \lambda_i\, u_a u_b,
\end{equation}
where $a_i$ describes the isotropic magnetic response of substate $i$, and $\lambda_i$ 
represents its uniaxial magnetic anisotropy. If $\theta$ is the angle between the applied field $\mathbf{B} = B\hat{\mathbf{b}}$ 
and the channel axis $\mathbf{u}$, with $\cos\theta = \hat{\mathbf{b}}\cdot\mathbf{u}$, the magnetic contribution to the free energy of substate $i$ becomes
\begin{equation}
\Delta\mathcal{F}_i(B,\theta) = -B^2 \big(a_i + \lambda_i \cos^2\theta\big) + \mathcal{O}(B^4).
\label{eq:dFi}
\end{equation}

Substituting Eq.~\eqref{eq:dFi} into the fluctuation--response relation~\eqref{eq:MasterResponse} 
gives the main analytical expression for the magnetic modulation of the current,
\begin{equation}
\Delta I(B, \theta, T, V) = \beta B^2 \big(C_{Ia}(T,V) + C_{I\lambda}(T,V)\cos^2\theta\big) + \mathcal{O}(B^4),
\label{eq:FinalPrediction}
\end{equation}
where the zero-field equilibrium covariances are defined as
\begin{equation}
C_{I\gamma}(T,V) = \langle I\gamma \rangle - \langle I \rangle \langle \gamma \rangle, \qquad \gamma \in \{a, \lambda\}.
\label{eq:covariances}
\end{equation}

A comparison between parallel ($\theta = 0$) and perpendicular ($\theta = \pi/2$) 
field orientations removes the isotropic contribution $C_{Ia}$,
\begin{equation}
\Delta I_{\parallel} - \Delta I_{\perp} \equiv \Delta I(B, 0) - \Delta I(B, \pi/2) = \beta B^2 C_{I\lambda} + \mathcal{O}(B^4),
\label{eq:AnisotropicResponse}
\end{equation}
so that the orientational contribution is determined by $C_{I\lambda}$.

Based on the quadratic field dependence required by time-reversal symmetry, we next examine the response of the channel using the \emph{magnetic sensitivity parameter} 
$S_B(T,B,\theta)$. This parameter is defined as the derivative of the stationary mean current $\langle I \rangle_{T,B,\theta} = \sum_{i=1}^{N} I_i\, p_i(T,B,\theta)$ with respect to 
the magnetic flux density $B$, at fixed temperature and field orientation:
\begin{equation}
    S_B(T,B,\theta) \equiv \left(\frac{\partial \langle I \rangle_{T,B,\theta}}{\partial B}\right)_{T,\theta} = -\beta \sum_{i=1}^{N} I_i \,p_i(T,B,\theta) \left( \frac{\partial \Delta \mathcal{F}_i}{\partial B} - \left\langle \frac{\partial \Delta \mathcal{F}}{\partial B} \right\rangle_{T,B,\theta} \right).
    \label{eq:SB_exact}
\end{equation}
Using $\partial \Delta \mathcal{F}_i / \partial B = -2B(a_i + \lambda_i \cos^2\theta)$, Eq.~\eqref{eq:SB_exact} gives the following fluctuation--sensitivity relation. The covariances in this expression are calculated using the field-dependent distribution $p_i(T,B,\theta)$, and the relation holds to the order retained in $\Delta\mathcal{F}_i$:
\begin{equation}
    S_B(T,B,\theta) = 2\beta B \left[ \mathrm{Cov}_{T,B,\theta}(I, a) + \mathrm{Cov}_{T,B,\theta}(I, \lambda)\cos^2\theta \right].
    \label{eq:SB_nonpert}
\end{equation}
In the weak-to-moderate field regime, where $\beta |\Delta \mathcal{F}_i| \ll 1$, the field-dependent covariances can be approximated by their zero-field equilibrium values, $C_{Ia}(T)$ and $C_{I\lambda}(T)$. This gives the linear sensitivity relation:
\begin{equation}
    S_B(T,B,\theta) = \frac{\partial \Delta I(T,B,\theta)}{\partial B} = 2\beta B \left[ C_{Ia}(T) + C_{I\lambda}(T)\cos^2\theta \right] + \mathcal{O}(\beta^2 B^3).
    \label{eq:SB_linear}
\end{equation}

In addition to the magnetic response, the temperature dependence of the pore current is also important because conformational gating in biological ion channels and functionalized nanopores is thermally activated. We therefore consider the sensitivity of the current to temperature changes as a second response quantity. The unperturbed conformational free energy of substate $i$ is written in terms of its internal energy $E_i$ and conformational entropy $\mathcal{S}_i$ as $\mathcal{F}_i^{(0)}(T) = E_i - T \mathcal{S}_i$. In the presence of a magnetic field, the corresponding effective internal energy is $E_{\mathrm{eff},i}(B,\theta) = E_i + \Delta \mathcal{F}_i(B,\theta)$. The equilibrium occupation probability of state $i$ is then
\begin{equation}
    p_i(T,B,\theta) = \frac{\exp\!\left[-\beta E_{\mathrm{eff},i}(B,\theta) + \mathcal{S}_i/k_{\mathrm{B}}\right]}{\sum_{j=1}^{N} \exp\!\left[-\beta E_{\mathrm{eff},j}(B,\theta) + \mathcal{S}_j/k_{\mathrm{B}}\right]}.
    \label{eq:P_energy_entropy}
\end{equation}
We define the \emph{thermal sensitivity parameter} $S_T(T,B,\theta)$ as the derivative of the stationary current with respect to temperature at fixed $(B,\theta)$. Assuming that the state currents $I_i$, internal energies $E_i$, conformational entropies $\mathcal{S}_i$, and magnetic free-energy contributions $\Delta \mathcal{F}_i$ are independent of temperature, this sensitivity can be expressed as
\begin{equation}
    S_T(T,B,\theta) \equiv \left(\frac{\partial \langle I \rangle_{T,B,\theta}}{\partial T}\right)_{B,\theta} = \frac{1}{k_{\mathrm{B}}T^2} \, \mathrm{Cov}_{T,B,\theta}\!\left(I, E_{\mathrm{eff}}\right),
    \label{eq:ST_exact}
\end{equation}
where $\mathrm{Cov}_{T,B,\theta}(I, E_{\mathrm{eff}}) = \langle I\,E_{\mathrm{eff}} \rangle_{T,B,\theta} - \langle I \rangle_{T,B,\theta}\langle E_{\mathrm{eff}} \rangle_{T,B,\theta}$ is the equilibrium cross-covariance between the state current and the effective internal energy. This relation shows that the thermal sensitivity is determined by the correlation between the ionic current and the effective energy of the conformational states. In particular, states with higher currents contribute differently to the temperature response depending on whether their effective energies are above or below the ensemble average.

\section{Microscopic Diamagnetic Estimate}
\label{sec:micro}

To relate the phenomenological coefficients $\lambda_i$ to the macromolecular structure of the channel, 
and to obtain realistic order-of-magnitude estimates, we connect the uniaxial free-energy 
shift with diamagnetic anisotropy~\cite{Worcester1978, Maret1986, Pauling1979}. 
In SI units, a structural domain with effective volume $v$ and dimensionless volume magnetic susceptibility tensor 
$\boldsymbol{\chi}$ has the following magnetostatic free-energy shift
\begin{equation}
\Delta \mathcal{F} = -\frac{v}{2\mu_0}\, \mathbf{B}\cdot\boldsymbol{\chi}\cdot\mathbf{B},
\label{eq:micro_free_energy}
\end{equation}
where $\mu_0$ is the vacuum permeability. For a structural domain that is symmetric around the channel axis 
$\mathbf{u}$, the susceptibility tensor can be written as
\begin{equation}
\boldsymbol{\chi} = \chi_\perp \mathbf{1} + \Delta\chi\, \mathbf{u}\otimes\mathbf{u}, \qquad \Delta\chi = \chi_\parallel - \chi_\perp,
\end{equation}
which gives
\begin{equation}
\Delta \mathcal{F} = -\frac{v B^2}{2\mu_0} \big(\chi_\perp + \Delta\chi \cos^2\theta\big).
\end{equation}
By comparing this result with Eq.~\eqref{eq:dFi}, the anisotropic response parameter can be identified as
\begin{equation}
\lambda = \frac{v\,\Delta\chi}{2\mu_0}.
\label{eq:lambda_micro}
\end{equation}

In transmembrane ion channels, the diamagnetic anisotropy mainly originates from the cumulative alignment of peptide bonds in $\alpha$-helical segments, including pore-lining S6/M2 helices and voltage-sensing S4 helices~\cite{Worcester1978, Torbet1981}. During conformational gating, processes such as helix tilting, splaying, or reorganization of the secondary structure can change the projection of these anisotropic structural motifs onto the pore axis $\mathbf{u}$~\cite{Delemotte2015, Goychuk2015}. At a coarse-grained level, we describe substate $i$ by an effective alignment index $n_i$, which represents the net amount of anisotropic structural units aligned along the channel axis:
\begin{equation}
\lambda_i = \lambda_0\, n_i, \qquad \lambda_0 = \frac{\Delta\chi_{\mathrm{eff}}\, v_0}{2\mu_0},
\label{eq:lambda_n}
\end{equation}
where $v_0$ and $\Delta\chi_{\mathrm{eff}}$ are the effective volume and axial susceptibility anisotropy of an elementary structural motif, for example, an aligned peptide group or a helical turn. The quantity $n_i$ does not necessarily represent an integer number of rigid structural blocks. Instead, it can be regarded as a continuous order parameter that describes the net orientational alignment, $n_i \sim \sum_{k} \frac{1}{2}(3\cos^2\alpha_k^{(i)} - 1)$, of $N_{\mathrm{tot}}$ anisotropic constituent units. Here, $\alpha_k^{(i)}$ is the angle between the $k$th structural unit and the channel axis $\mathbf{u}$ in substate $i$.

Using the scaling relation~\eqref{eq:lambda_n}, the anisotropic covariance can be separated into a microscopic material parameter $\lambda_0$ and a structural--transport covariance $C_{In}$:
\begin{equation}
C_{I\lambda} = \lambda_0\, C_{In}, \qquad C_{In} = \langle I n \rangle - \langle I \rangle \langle n \rangle,
\label{eq:CIn}
\end{equation}
and the orientation-dependent contribution to the current response is therefore
\begin{equation}
\Delta I_{\mathrm{aniso}}(B, \theta, T, V) = \beta B^2 \lambda_0\, C_{In}(T,V)\cos^2\theta + \mathcal{O}(B^4).
\label{eq:aniso_prediction}
\end{equation}
Equation~\eqref{eq:aniso_prediction} describes the physical origin of directional magnetosensitivity in the present multistate model. An orientational response appears when conformational transitions associated with a more open pore, $I_i > \langle I \rangle$, are systematically correlated, $C_{In} > 0$, or anticorrelated, $C_{In} < 0$, with the axial alignment of diamagnetically anisotropic protein motifs, $n_i \gtrless \langle n \rangle$. Thus, both a structural variation in magnetic anisotropy and its correlation with the state-dependent ionic current are required for a nonzero directional response. At the same time, since $\lambda_0 \sim 10^{-29}~\mathrm{J\,T^{-2}}$ per peptide or helical unit (see Eq.~\eqref{eq:lambda0_value} below), Eq.~\eqref{eq:aniso_prediction}, together with the Cauchy--Schwarz inequality, gives an upper bound on single-channel magnetosensitivity in weak magnetic fields. We examine this constraint explicitly in Section~\ref{sec:observability}.

Within the structural order-parameter description of Eq.~\eqref{eq:lambda_n}, the anisotropic coupling of a given state is written as $\lambda_i = \lambda_0 n_i$, where $\lambda_0 = 
\Delta\chi_{\mathrm{eff}} v_0 / (2\mu_0)$ is the diamagnetic energy coefficient associated with one effective structural unit, such as a peptide bond or an $\alpha$-helical residue. 
Based on high-field magnetic birefringence and nuclear magnetic resonance measurements for oriented polypeptides and transmembrane 
helices~\cite{Worcester1978,Pauling1979,Torbet1981,Maret1986}, the volume susceptibility anisotropy of $\alpha$-helical protein segments is estimated as $|\Delta\chi_{\mathrm{eff}}| 
\sim 10^{-6}$ in SI units. The effective molecular volume per residue is of the order of $v_0 \sim 10^{-28}~\text{m}^3$. Using $\mu_0 = 4\pi \times 10^{-7}~\text{H\,m}^{-1}$, the 
resulting elementary coupling constant is
\begin{equation}
    |\lambda_0| = \frac{|\Delta\chi_{\mathrm{eff}}|\,v_0}{2\mu_0} \simeq 4.0 \times 10^{-29}~\text{J\,T}^{-2}.
    \label{eq:lambda0_value}
\end{equation}
For a macromolecular ion channel or nanopore containing $n_i \sim 10^2\text{--}10^3$ aligned anisotropic peptide units that undergo concerted tilting during gating, the corresponding d
difference in conformational susceptibility can range over $|\Delta \lambda| = |\lambda_0|\,\Delta n \sim 4 \times 10^{-27}$ to $4 \times 10^{-26}~\text{J\,T}^{-2}$.

{

\section{Three-State Landscape and Observable Predictions}
\label{sec:magnetic}

The statistical-mechanical formalism developed in Sections~\ref{sec:model}--\ref{sec:micro} provides testable predictions for the response of the stationary current in a multistate ion channel or nanopore to external magnetic fields and temperature variations. In this section, we bring together the main physical selection rules described by the sensitivity parameters $S_B$ and $S_T$, examine the condition for sign reversal, and study their temperature dependence in the physiological regime using an explicit three-state conformational landscape.

\subsection{Sign-reversal boundary and magnetic scaling rules}
\label{subsec:sign_symmetry}

Together with the microscopic order-parameter relation $\lambda_i = \lambda_0 n_i$ given in Eq.~\eqref{eq:lambda_n}, the response relations~\eqref{eq:FinalPrediction} and \eqref{eq:SB_linear}--\eqref{eq:ST_exact} lead to three observable properties of the magnetic response: \\
\textit{(i) Linear scaling of sensitivity with magnetic flux density ($S_B \propto B$):} 
Since the diamagnetic free-energy shift is quadratic in $B$ ($\Delta I \propto B^2$), the differential sensitivity parameter $S_B$ is zero at $B = 0$ and increases linearly with the 
magnetic field. Its slope, $\partial S_B / \partial B = 2\beta(C_{Ia} + C_{I\lambda}\cos^2\theta)$, directly reflects the equilibrium covariance between the state-dependent current and 
the magnetic response.\\
\textit{(ii) Quadrupolar angular dependence ($\cos^2\theta$):} Rotation of the magnetic field changes $S_B(T,B,\theta)$ according to a uniaxial $\cos^2\theta$ dependence. By comparing the parallel and perpendicular field orientations using Eq.~\eqref{eq:SB_linear}, and defining $\Delta S_B(T,B) \equiv S_B(T,B,0) - S_B(T,B,\pi/2)$, one obtains $\Delta S_B(T,B) = 2\beta B \lambda_0 C_{In}(T)$. This comparison makes it possible, in principle, to separate the structural anisotropy covariance $C_{I\lambda} = \lambda_0 C_{In}$ from an isotropic background contribution $C_{Ia}$.\\
\textit{(iii) Thermodynamic sign-reversal boundary ($C_{I\lambda} = 0$):} The sign of both the current shift $\Delta I$ and the sensitivity parameter $S_B$ is determined by the sign of the equilibrium covariance. In particular, current enhancement ($\Delta I > 0$, $S_B > 0$) occurs when $C_{Ia} + C_{I\lambda}\cos^2\theta > 0$. This means that the more conducting substates have, on average, a more positive effective magnetic response coefficient $a_i + \lambda_i\cos^2\theta$ than the less conducting substates. Conversely, current suppression ($\Delta I < 0$, $S_B < 0$) occurs when $C_{Ia} + C_{I\lambda}\cos^2\theta < 0$. For the purely anisotropic response, the locus
    \begin{equation}
        C_{I\lambda}(T^*) = \langle I\,\lambda \rangle_{0} - \langle I \rangle_{0}\langle \lambda \rangle_{0} = 0
        \label{eq:prediction1}
    \end{equation}
    defines the sign-reversal boundary of the anisotropic response. This condition is model independent in the sense that it only depends on the sign of a covariance. When $C_{I\lambda} > 0$, an axial magnetic field enhances the current relative to a transverse field ($\Delta I_\parallel - \Delta I_\perp > 0$, $\Delta S_B > 0$), whereas for $C_{I\lambda} < 0$, the corresponding response is suppressive ($\Delta I_\parallel - \Delta I_\perp < 0$, $\Delta S_B < 0$).
\begin{figure}[htbp]
    \centering
    \includegraphics[width=0.72\columnwidth]{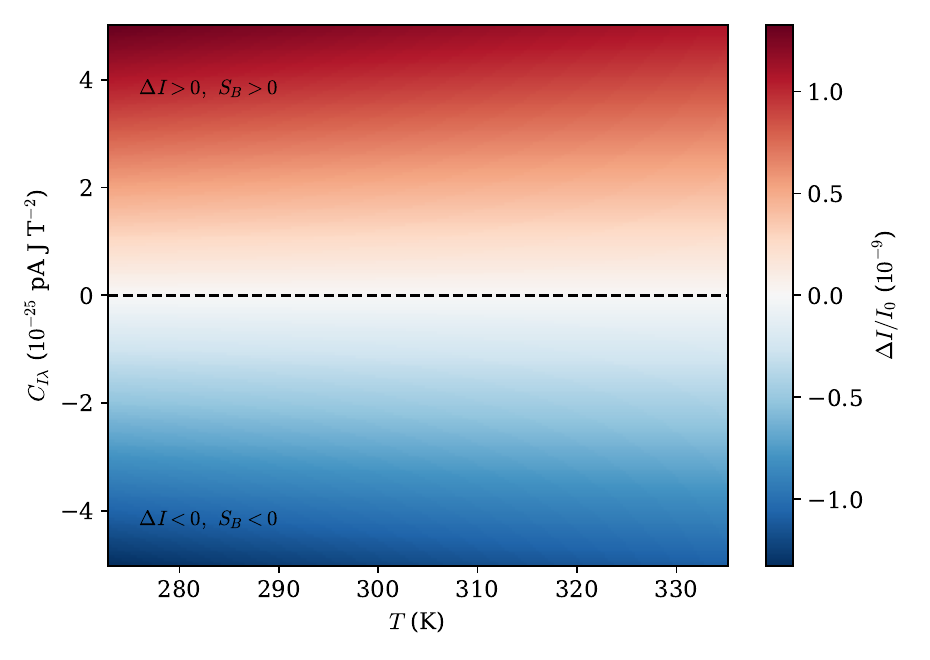}
    \caption{Sign map of the normalized leading-order magnetic current response $\Delta I / I_0$ at a representative field strength $B = 10~\text{mT}$ across the $(T, C_{I\lambda})$ plane (for negligible isotropic background, $C_{Ia} \simeq 0$). The vertical dashed locus $C_{I\lambda} = 0$ marks the thermodynamic sign-reversal boundary separating positive sensitivity ($\Delta I > 0$, $S_B > 0$) and negative sensitivity ($\Delta I < 0$, $S_B < 0$) regimes.}
    \label{fig:heatmap}
\end{figure}
Figure~\ref{fig:heatmap} illustrates this sign-reversal boundary in the $(T, C_{I\lambda})$ plane for $C_{Ia} \simeq 0$. The vertical threshold $C_{I\lambda} = 0$ separates the response into positive- and negative-sensitivity regions. At a fixed value of the covariance, increasing the temperature reduces the response amplitude through the prefactor $\beta = (k_{\mathrm{B}}T)^{-1}$.

\subsection{Three-state model: magnetic and thermal sensitivity}
\label{subsec:three_state_results}

To illustrate the quantitative behavior of the sensitivity parameters $S_B$ and $S_T$, we evaluate Eqs.~\eqref{eq:SB_nonpert} and \eqref{eq:ST_exact} for an explicit three-state conformational channel ($N=3$). The model contains a closed ground state ($i=1$, $I_1 = 0$), an intermediate pre-open or sub-conducting state ($i=2$, $I_2 > 0$), and a fully open conducting state ($i=3$, $I_3 > I_2$). Thermally activated transitions from the closed state to the higher-energy and higher-entropy conducting states ($E_1 < E_2 < E_3$, $\mathcal{S}_1 < \mathcal{S}_2 < \mathcal{S}_3$) are associated with concerted reorganization of the secondary structure, for example, $\alpha$-helical tilting. We therefore assign different diamagnetic anisotropy coefficients, $\lambda_1 < \lambda_2 < \lambda_3$, to the three states.

The parameters used in the following figures are listed in Table~\ref{tab:params}. They are based on the heat-activated channel TRPV1 as a representative thermosensitive channel. An activation enthalpy of about $90~\text{kcal/mol}$ with half-activation near $42^{\circ}\text{C}$ is reported in \cite{Yao2011}, while a unitary slope conductance of $80~\text{pS}$ at $+60~\text{mV}$ is reported in \cite{Oh1996}, corresponding to $I_3 \simeq 4.8~\text{pA}$. For the intermediate state, we use an enthalpy of $20~\text{kcal/mol}$, which is within the typical range for ion-channel gating \cite{Hille2001, Yao2010, Yao2011}. The sub-conducting current $I_2$, the closed--intermediate midpoint ($295~\text{K}$), and the alignment indices $n_i$ are model assumptions rather than measured quantities. The entropy of the open state is chosen such that $p_3 = 1/2$ at $T = 315~\text{K}$. As assumed in Sec.~\ref{sec:quadratic}, the temperature dependence of the unitary currents is neglected.

\begin{table}[htbp]
    \centering
    \caption{Parameters of the three-state model used in Figs.~\ref{fig:thermal_sensitivity} and \ref{fig:2D_landscape} (TRPV1-based; see text for sources and assumptions). Energies and entropies are given relative to the closed state. The anisotropy coefficients are $\lambda_i = \lambda_0 n_i$ with the baseline value $\lambda_0 = 4\times 10^{-29}~\text{J\,T}^{-2}$ of Eq.~\eqref{eq:lambda0_value}; $a_i = 0$ and $\theta = 0$.}
    \label{tab:params}
    \begin{tabular}{c c c c c}
        \hline
        State $i$ & $I_i$ (pA) & $E_i$ (kJ\,mol$^{-1}$) & $\mathcal{S}_i$ (J\,mol$^{-1}$\,K$^{-1}$) & $n_i$ \\
        \hline
        1 (closed)       & 0   & 0    & 0   & 0   \\
        2 (intermediate) & 2.0 & 83.7 & 284 & 100 \\
        3 (open)         & 4.8 & 376.6 & 1214 & 300 \\
        \hline
    \end{tabular}
\end{table}

\begin{figure}[htbp]
    \centering
    \includegraphics[width=0.48\columnwidth]{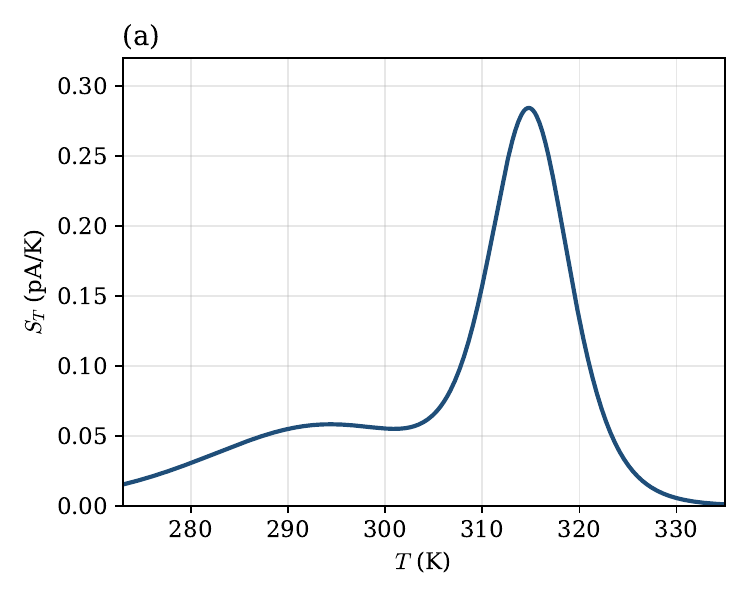}\hfill
    \includegraphics[width=0.48\columnwidth]{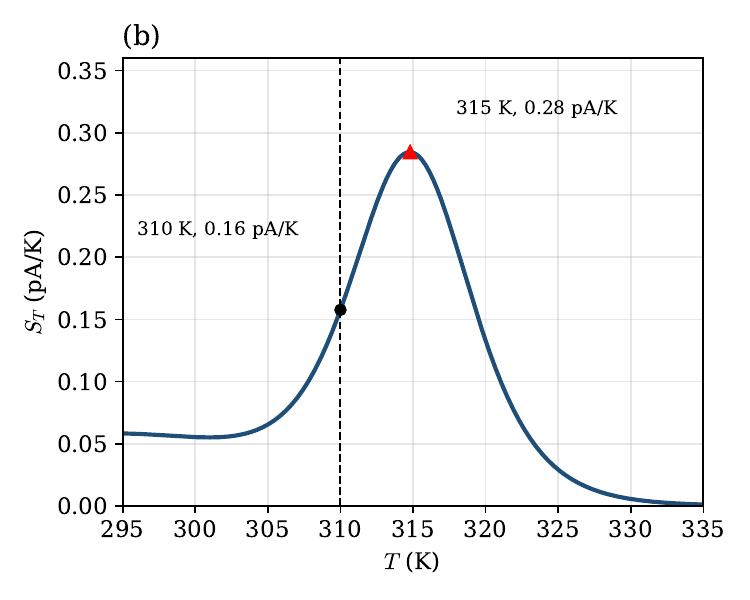}
    \caption{Thermal sensitivity parameter $S_T = \partial \langle I \rangle / \partial T$ ($\text{pA/K}$) as a function of temperature across the biological window $T \in [273, 335]~\text{K}$. \textbf{(a) Left panel:} Broad-range profile of the sensitivity of the channel to temperature variations, showing the transition from a quiescent low-temperature state ($T \lesssim 305~\text{K}$) to rapid thermal activation above $305~\text{K}$. \textbf{(b) Right panel:} Response of the channel in the physiological range, highlighting the steep rise at the physiological temperature $T = 310~\text{K}$ (vertical dashed line, $S_T \simeq 0.16~\text{pA/K}$) and the sensitivity maximum at $T_{\max} \approx 315~\text{K}$ ($S_T \simeq 0.28~\text{pA/K}$) prior to high-temperature saturation.}
    \label{fig:thermal_sensitivity}
\end{figure}

Figure~\ref{fig:thermal_sensitivity} shows the temperature dependence of the thermal sensitivity parameter $S_T(T)$ over the range $T \in [273, 335]~\text{K}$. According to Eq.~\eqref{eq:ST_exact}, three different operating regimes can be identified:\\
 \textit{(i) Quiescent low-temperature regime ($273~\text{K} \le T \lesssim 305~\text{K}$):} In this temperature range, thermal agitation is weak compared with the energy gap to the open state. The population is therefore mainly shared between the closed and intermediate states, while the open-state population remains very small ($p_3 < 0.02$). As a result, the energy--current covariance $C_{IE}(T)$ is small, and the thermal sensitivity remains low, with $S_T \lesssim 0.07~\text{pA/K}$ (Fig.~\ref{fig:thermal_sensitivity}, left panel).\\
 \textit{(ii) Physiological activation regime ($305~\text{K} \lesssim T \lesssim 315~\text{K}$):} As the temperature enters the physiological range, thermal excitation increasingly populates the open conducting state. Since $\mathrm{Cov}(I,E)$ is determined by products of competing state probabilities $p_i p_j$, $S_T(T)$ increases rapidly in this region. The normal physiological temperature $T = 310~\text{K}$, marked by the vertical dashed line in the right panel of Fig.~\ref{fig:thermal_sensitivity}, corresponds to $S_T \simeq 0.16~\text{pA/K}$ and lies in the transition region with a relatively large slope, just below the sensitivity maximum.\\
\textit{(iii) Sensitivity peak and high-temperature saturation ($T \gtrsim 315~\text{K}$):} The thermal sensitivity reaches its maximum value of $S_T \simeq 0.28~\text{pA/K}$ at $T_{\max} \approx 315~\text{K}$ ($\approx 42^{\circ}\text{C}$, where $\partial S_T / \partial T = 0$). This maximum is associated with the coexistence of intermediate and open conformations ($p_2 \approx p_3 \approx 0.5$). At higher temperatures, the open state becomes increasingly saturated ($p_3 \to 1$), the conformational variance decreases, and $S_T$ drops rapidly, reaching approximately $0.03~\text{pA/K}$ at $325~\text{K}$.

\begin{figure}[htbp]
    \centering
    \includegraphics[width=0.92\columnwidth]{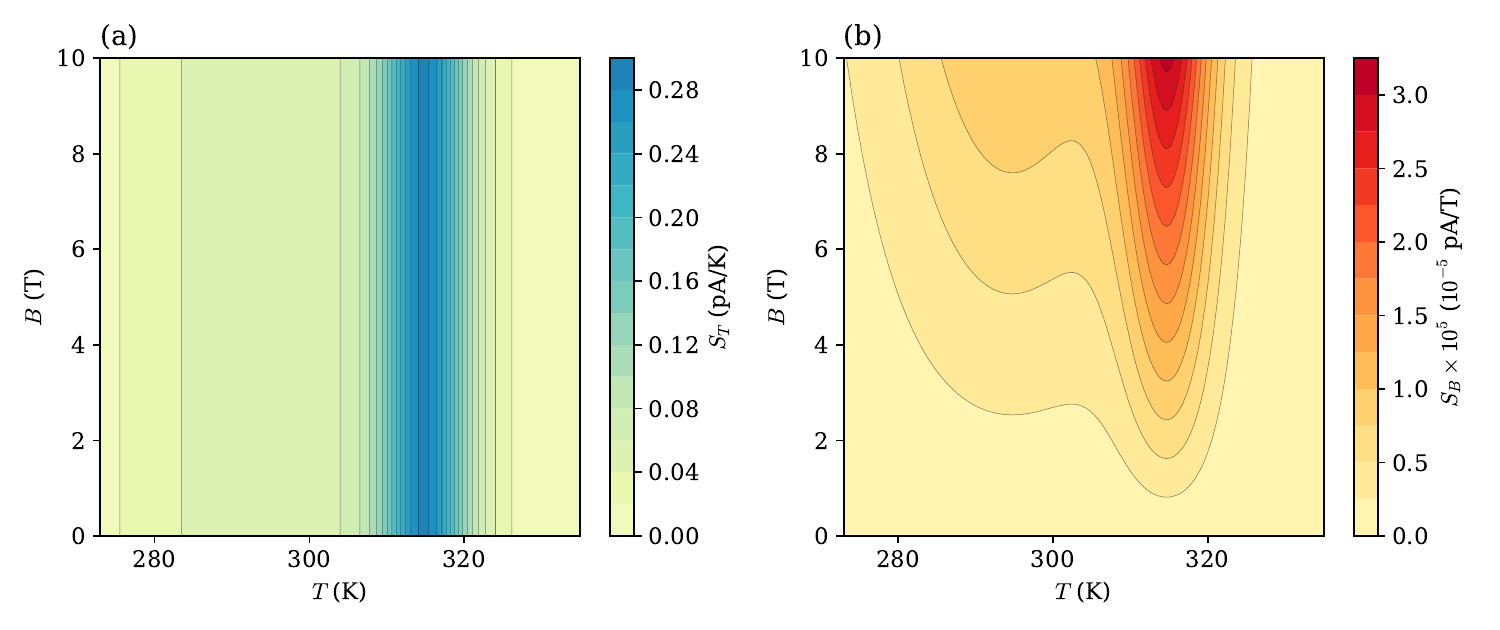}
    \caption{Two-dimensional thermo-magnetic sensitivity landscapes across the $(B,T)$ control plane ($B \in [0, 10]~\text{T}$, $T \in [273, 335]~\text{K}$) for the three-state channel model. \textbf{(a) Left panel:} Contour map of the thermal responsivity (thermal sensitivity parameter) $S_T = \partial \langle I \rangle / \partial T$ ($\text{pA/K}$), featuring a pronounced ridge, extending along $B$, peaking at $S_T \simeq 0.28~\text{pA/K}$ around $T \approx 315~\text{K}$. \textbf{(b) Right panel:} Contour map of the scaled magnetic sensitivity parameter $S_B \times 10^5 = (\partial \langle I \rangle / \partial B) \times 10^5$ (in units of $10^{-5}~\text{pA/T}$), showing how operating near the thermal activation peak ($\approx 315~\text{K}$) enhances the magnetic sensitivity.}
    \label{fig:2D_landscape}
\end{figure}

To show the response in the two-dimensional parameter space, Fig.~\ref{fig:2D_landscape} presents contour maps of the thermal responsivity $S_T = \partial \langle I \rangle / \partial T$ (left panel) and the scaled magnetic sensitivity parameter $S_B \times 10^5 = (\partial \langle I \rangle / \partial B) \times 10^5$ (right panel) in the $(B,T)$ plane. The quantity $S_T(B,T)$ forms a ridge that extends along the $B$ direction and is centered near $T \approx 315~\text{K}$. In this range, $S_T$ is essentially independent of $B$, so Fig.~\ref{fig:thermal_sensitivity} can effectively be viewed as representing the zero-field thermal sensitivity. In contrast, the magnetic sensitivity $S_B(B,T)$ shown in the right panel increases linearly with $B$ at a fixed temperature, in agreement with Eq.~\eqref{eq:SB_linear}, and becomes significantly larger near the thermal activation peak.

For example, at $B = 10~\text{T}$, $S_B$ increases from approximately $3\times 10^{-6}~\text{pA/T}$ at $275~\text{K}$ to about $2\times 10^{-5}~\text{pA/T}$ at $310~\text{K}$, reaching a maximum of approximately $3\times 10^{-5}~\text{pA/T}$ ($0.03~\text{fA/T}$) at $315~\text{K}$. This corresponds to an increase of approximately one order of magnitude. The linear dependence indicates that the leading-order covariance relation remains valid, since for a single channel $\beta|\Delta\mathcal{F}_i| \lesssim 3\times 10^{-4}$ throughout this range for the parameters listed in Table~\ref{tab:params}. The calculated values are well below the baseline bound obtained from Eq.~\eqref{eq:SB_bound} after rescaling to $B = 10~\text{T}$, which is approximately $2\times 10^{-3}~\text{pA/T}$. This is expected because the variations of $n_i$ and $I_i$ in the three-state model are smaller than the values $\sigma_n \sim 10^3$ and $\sigma_I \sim 10~\text{pA}$ assumed in deriving that bound. From a physical point of view, $S_B \propto \beta C_{I\lambda}(T)$ is small at low temperatures because the open-state population is negligible. At high temperatures, it is also reduced because saturation of the open state removes much of the conformational variance. The maximum magnetic sensitivity therefore appears in the temperature region where the intermediate and open conformations coexist.

\section{Bounds and Observability}
\label{sec:observability}

The covariance relations developed in Sections~\ref{sec:model}--\ref{sec:magnetic} determine the functional dependence and the sign criteria of $\Delta I$, $S_B$, and $S_T$. However, in order to assess whether these effects can be detected experimentally, quantitative bounds based on molecular diamagnetism are also needed. In this section, we derive upper bounds based on the Cauchy--Schwarz inequality for the relative current change and for the differential magnetic sensitivity $\Delta S_B$ within the present framework. We then evaluate these bounds over a range extending from weak millitesla fields to strong fields in the Tesla range. We also discuss how cooperative clustering of channels may provide a possible connection between the very small energetics of a single molecule and macroscopic electrophysiological measurements.

\subsection{Cauchy--Schwarz upper bounds on sensitivity}
\label{subsec:CS_bound}

The Cauchy--Schwarz inequality can be applied to the equilibrium covariance $C_{I\lambda} = \mathrm{Cov}_0(I,\lambda)$. This provides a bound on the magnitude of the coupling in terms of the equilibrium standard deviations of the state current, $\sigma_I = \sqrt{\langle I^2 \rangle_0 - \langle I \rangle_0^2}$, and the structural order parameter, $\sigma_n = \sqrt{\langle n^2 \rangle_0 - \langle n \rangle_0^2}$:
\begin{equation}
    |C_{I\lambda}(T)| \le \sigma_I(T) \, \sigma_\lambda(T) = |\lambda_0| \, \sigma_I(T) \, \sigma_n(T).
    \label{eq:CS_inequality}
\end{equation}
By substituting Eq.~\eqref{eq:CS_inequality} into Eq.~\eqref{eq:aniso_prediction}, and also into the expression for $\Delta S_B$ given in Sec.~\ref{subsec:sign_symmetry}, upper limits can be obtained for the relative anisotropic current change and for the differential magnetic sensitivity $\Delta S_B$:
\begin{align}
    \frac{|\Delta I_{\mathrm{aniso}}|}{|\langle I \rangle_0|} &\le \beta B^2 |\lambda_0| \, \sigma_n \left( \frac{\sigma_I}{|\langle I \rangle_0|} \right) \cos^2\theta, \label{eq:relchange} \\
    |\Delta S_B(T,B)| &\le 2\beta B |\lambda_0| \, \sigma_n \, \sigma_I, \label{eq:SB_bound}
\end{align}

At physiological temperature ($T = 310~\text{K}$, $k_{\mathrm{B}}T \simeq 4.28 \times 10^{-21}~\text{J}$), the dimensionless factor associated with a single structural unit at a weak laboratory field of $B = 10~\text{mT}$ is $\beta B^2 |\lambda_0| \simeq 9.3 \times 10^{-13}$. Even if a relatively large conformational rearrangement involving $\sigma_n \sim 10^3$ aligned units is considered, together with an order-unity relative current fluctuation ($\sigma_I / |\langle I \rangle_0| \sim 1$, with a typical single-channel current standard deviation $\sigma_I \sim 10~\text{pA}$), Eqs.~\eqref{eq:relchange} and \eqref{eq:SB_bound} give
\begin{equation}
    \left. \frac{|\Delta I_{\mathrm{aniso}}|}{|\langle I \rangle_0|} \right|_{B=10\,\text{mT}} \lesssim 10^{-9} \cos^2\theta, \qquad \left. |\Delta S_B| \right|_{B=10\,\text{mT}} \lesssim 2 \times 10^{-6}~\text{pA/T}.
    \label{eq:upperbound}
\end{equation}
This estimate helps explain why direct diamagnetic shifts of the gating of an isolated single channel at millitesla fields are not expected to be resolved above thermal Johnson--Nyquist noise and shot noise.

\subsection{Pathways to experimental observability}
\label{subsec:observability_pathways}

At the same time, Eqs.~\eqref{eq:relchange} and \eqref{eq:SB_bound} suggest several possible physical mechanisms that could increase the response and bring the sensitivity closer to an experimentally measurable range:\\
\textit{- High-field scaling and aromatic/liquid-crystalline functionalization ($B \sim 1\text{--}30~\text{T}$):} The present model predicts $\Delta I \propto B^2$ and $S_B \propto B$. Therefore, increasing the magnetic flux density from $10~\text{mT}$ to the range of strong superconducting magnets ($B = 10~\text{T}$, as considered in Fig.~\ref{fig:2D_landscape}, and up to $30~\text{T}$ in high-field facilities) increases $\Delta I$ by a factor of $10^6\text{--}10^7$ and the differential sensitivity $S_B$ by $10^3\text{--}3 \times 10^3$. In addition, biomimetic solid-state nanopores functionalized with stacked aromatic rings, carbon nanotube porins, or liquid-crystalline lipid bilayers may have an effective susceptibility anisotropy $|\Delta\chi_{\mathrm{eff}}|$ that is one or two orders of magnitude larger than that of simple aliphatic helices. This is used here as an estimate. Since the response depends linearly on $\Delta\chi_{\mathrm{eff}}$, such an increase would raise the single-pore sensitivity shown in Fig.~\ref{fig:2D_landscape}(b) from approximately $3\times 10^{-5}~\text{pA/T}$ to about $3\times 10^{-4}\text{--}3\times 10^{-3}~\text{pA/T}$ at $10~\text{T}$. \\
\textit {- Cooperative amplification in membrane channel clusters ($N_{\mathrm{ch}}$):} In both native biological membranes and synthetic sensor arrays, ion channels can form allosterically coupled microdomains or lipid-raft clusters containing $N_{\mathrm{ch}}$ channels that may gate cooperatively~\cite{Hille2001}. If a cluster of $N_{\mathrm{ch}}$ coupled channels, or an oriented phospholipid domain containing mechanosensitive or voltage-gated pores, undergoes a concerted conformational transition, the effective correlated structural fluctuation scales in the fully correlated limit as $\sigma_{n,\mathrm{coop}} = N_{\mathrm{ch}}\sigma_n$. The total current fluctuation of the cluster similarly scales as $\sigma_{I,\mathrm{coop}} = N_{\mathrm{ch}}\sigma_I$. According to Eq.~\eqref{eq:SB_bound}, the upper limit for the collective cluster sensitivity parameter $S_B^{(\mathrm{coop})}$ then has a quadratic dependence on the size of the cooperative domain:
    \begin{equation}
        S_B^{(\mathrm{coop})}(T,B) \sim N_{\mathrm{ch}}^2 \, S_B^{(\mathrm{single})}(T,B),
        \label{eq:coop_scaling}
    \end{equation}
    For macroscopic whole-cell or multi-pore measurements involving $M$ independent domains, the signal-to-noise ratio can also improve as $\sqrt{M}$. Thus, even cooperative clusters of moderate size, $N_{\mathrm{ch}} \sim 10^2\text{--}10^3$, operating near their physiological thermal peak ($T \approx 310\text{--}320~\text{K}$, Fig.~\ref{fig:thermal_sensitivity}), could increase the upper limit on $S_B^{(\mathrm{coop})}$ from approximately $3\times 10^{-5}~\text{pA/T}$ to values in the range $0.3\text{--}30~\text{pA/T}$ at $B = 10~\text{T}$. Since the linearized approximation requires $\beta|\Delta\mathcal{F}_i| \ll 1$, its validity needs to be checked again for large cooperative clusters and strong magnetic fields. For this reason, these values should be interpreted only as upper-limit estimates. \\
\textit{-Experimental protocol for measuring $S_T$, $S_B$, and $C_{I\lambda}$:} The predictions of the present model can be examined experimentally using a two-stage procedure. First, zero-field temperature sweeps over $T \in [273, 335]~\text{K}$ can be performed to obtain the thermal sensitivity curve $S_T(T)$ (Fig.~\ref{fig:thermal_sensitivity}) and to identify the physiological temperature range ($T \approx 310\text{--}320~\text{K}$) where the conformational variance is largest. In the second stage, the temperature can be stabilized near this peak and a low-frequency modulated magnetic field, $B(t) = B_0 \cos(\omega t)$, can be applied at the two orientations $\theta = 0$ and $\theta = \pi/2$. The quadratic response $\Delta I$, and consequently $S_B(T,B,\theta)$, can then be extracted at the second harmonic ($2\omega$) using lock-in detection. The simultaneous observation of the linear dependence $S_B \propto B$ (Fig.~\ref{fig:2D_landscape}b), the $\cos^2\theta$ quadrupolar symmetry, and the enhancement around the thermal peak (Fig.~\ref{fig:2D_landscape}) would provide a characteristic signature of covariance-driven diamagnetic gating.

Finally, it should be noted that the quasi-equilibrium Boltzmann distribution in Eq.~\eqref{eq:Boltzmann} assumes that conformational gating relaxations occur sufficiently rapidly compared with changes in the external parameters. It also assumes that the ionic flux through the open pore does not strongly modify the conformational free-energy landscape. A possible extension of the present fluctuation--sensitivity formalism, involving $S_B$ and $S_T$, would be to consider nonequilibrium driven Markov networks with nonzero probability currents. This represents a natural direction for future investigation.

}

\section{Discussion and Conclusion}
\label{sec:discussion}

In this work, we have developed a statistical-mechanical framework to investigate the thermo-magnetic modulation of steady-state ionic transport across a multistate channel system. Rather than invoking direct Lorentz forces on mobile ions or ad hoc modifications of single-state pore conductances, our approach treats the channel macromolecule as an ensemble of discrete conformational substates residing on a rugged free-energy landscape~\cite{Frauenfelder1991}. Within this paradigm, an external magnetic field acts as a thermodynamic control parameter that tilts the conformational energy landscape via the anisotropic diamagnetic susceptibility of ordered secondary-structure motifs, thereby redistributing the equilibrium Boltzmann populations and shifting the ensemble-averaged ionic current.

A primary theoretical outcome of this study is that the leading-order magnetic response of the mean current is governed by a linear fluctuation--response relation. Because time-reversal symmetry of the equilibrium conformational free energies forbids linear terms in $\mathbf{B}$, the lowest-order response scales quadratically with the field magnitude ($B^2$) and is determined by the equilibrium covariance between the state-dependent ionic currents and the state-resolved magnetic susceptibility coefficients. Specifically, the orientation-dependent modulation obeys a $\cos^2\theta$ angular law controlled by the structural covariance $C_{In} = \langle I n \rangle_0 - \langle I \rangle_0 \langle n \rangle_0$, where $n_i$ quantifies the net effective alignment of diamagnetically anisotropic structural units (such as peptide bonds in $\alpha$-helices) along the channel symmetry axis. Consequently, the sign, magnitude, and angular selectivity of the current modulation do not depend on the absolute magnetic susceptibility of the protein as a whole, but rather on the statistical cross-correlation between conformational geometry and state conductance across the active ensemble. In particular, the vanishing of $C_{In}$ defines a compensation condition under which the orientation-dependent part of the response vanishes despite nonzero state susceptibilities and nonzero state currents.

To connect this statistical formalism with experimentally measurable observables, we introduced two differential sensitivity parameters: the magnetic sensitivity $S_B(T,B,\theta) = \partial \langle I \rangle/\partial B$ and the thermal sensitivity $S_T(T,B,\theta) = \partial \langle I \rangle/\partial T$. While $S_B$ scales linearly with $B$ in the weak-field regime and maps directly onto the structural covariance $C_{In}$, the thermal sensitivity is governed by the covariance between the state current and the effective energy, Eq.~\eqref{eq:ST_exact}. In the three-state illustration, $S_T$ is non-monotonic in temperature, with a maximum near $T \approx 315~\text{K}$. The field-induced shift of $S_T$ obtained here contains only the explicit $\propto -T^{-2}$ prefactor; the additional temperature dependence of the covariances themselves is not analyzed.

From a quantitative standpoint, our order-of-magnitude analysis and Cauchy--Schwarz bounds establish a transparent physical benchmark for assessing claims of magnetoreception and magnetic field effects in biological membranes. For a single ion channel with a conformational fluctuation $\sigma_n \sim 10^3$ of aligned peptide units, the characteristic energy scale of the diamagnetic coupling per unit at millitesla fields ($\sim 10\text{ mT}$) at physiological temperature is $\beta B^2 |\lambda_0| \sim 10^{-12}$, yielding an upper bound on the fractional single-channel current modulation of $|\Delta I_{\mathrm{aniso}}|/\langle I \rangle_0 \lesssim 10^{-9}$. Measurable single-molecule modulations on the order of $10^{-2}$ via purely diamagnetic population shifting require either ultra-high static magnetic fields ($B \sim 20\text{--}30\text{ T}$) or large-scale cooperative coupling across supramolecular channel clusters and membrane domains. Within the present idealized model, this constraint implies that a robust response to weak (millitesla or geomagnetic) fields cannot arise from equilibrium diamagnetic tilting of isolated, non-interacting channel proteins, and would have to rely on collective effects, non-equilibrium kinetic amplification, or radical-pair/paramagnetic mechanisms.

Several theoretical limitations of the present model merit explicit consideration and point toward natural extensions for future research:\\
 \textit{Time-scale separation and non-equilibrium steady states (NESS):} Our derivation assumes a strict separation of time scales wherein conformational populations relax to a quasi-equilibrium Boltzmann distribution unaffected by the ionic flux. In many biological and synthetic nanopores, however, ion occupancy within the selectivity filter, electro-conformational coupling, and voltage-dependent gating transitions violate detailed balance. Extending this framework to a fully coupled Markovian master equation or Fokker--Planck network—where transition rates $W_{ij}(\mathbf{B}, V)$ break detailed balance—will allow the characterization of non-equilibrium covariance criteria and flux-driven sensitivity amplification. \\
 \textit{Cooperative allostery and ensemble coupling:} The current model treats a single, independent channel. Incorporating short-range mechanical coupling (via membrane lipid deformation) or long-range dipolar interactions among clustered channels within an Ising- or Potts-type lattice gas model could enhance the effective structural variance $\sigma_n^2$ near critical demixing or gating points, substantially amplifying both $S_B$ and $S_T$.\\
 \textit{Explicit free-energy parametrizations:} While we kept the unperturbed conformational free energies $\mathcal{F}_i^{(0)}$ general to preserve model-independent validity, combining the present fluctuation--response relations with molecular dynamics (MD) trajectories and enhanced-sampling free-energy surfaces will enable the direct computation of $C_{In}$ and $C_{IE}$ for specific voltage-gated or mechanosensitive channel families.

In summary, by bridging conformational landscape theory with linear fluctuation--response relations, this study provides an analytical baseline for understanding how external vector fields and thermal fluctuations jointly modulate transport in multistate nanoscale systems.

\section*{Acknowledgements}
This work is partially supported by the Iran National Science Foundation (INSF).

\bibliographystyle{unsrt}
\bibliography{mybib}

\end{document}